\documentclass[aps,prl,twocolumn,showpacs,superscriptaddress,preprintnumbers,amsmath,amssymb]{revtex4}
\usepackage{graphicx}
\usepackage{dcolumn}
\usepackage{bm}
\usepackage{color}
\usepackage{adjustbox}
\usepackage{multirow}
\usepackage{ulem}

\begin{document}

\title{Breakdown of the spectator concept in low-electron-energy resonant decay processes}

\author{\firstname{A.} \surname{Mhamdi}}
\affiliation{Institut f\"{u}r Physik und CINSaT, Universit\"{a}t Kassel, Heinrich-Plett-Str. 40, 34132 Kassel, Germany}

\author{\firstname{J.} \surname{Rist}}
\affiliation{Institut f\"{u}r Kernphysik, J. W. Goethe-Universit\"{a}t, Max-von-Laue-Strasse 1, 60438 Frankfurt am Main, Germany}

\author{\firstname{D.} \surname{Aslit\"urk}}
\affiliation{Institut f\"{u}r Kernphysik, J. W. Goethe-Universit\"{a}t, Max-von-Laue-Strasse 1, 60438 Frankfurt am Main, Germany}

\author{\firstname{M.} \surname{Weller}}
\affiliation{Institut f\"{u}r Kernphysik, J. W. Goethe-Universit\"{a}t, Max-von-Laue-Strasse 1, 60438 Frankfurt am Main, Germany}

\author{\firstname{N.} \surname{Melzer}}
\affiliation{Institut f\"{u}r Kernphysik, J. W. Goethe-Universit\"{a}t, Max-von-Laue-Strasse 1, 60438 Frankfurt am Main, Germany}

\author{\firstname{D.} \surname{Trabert}}
\affiliation{Institut f\"{u}r Kernphysik, J. W. Goethe-Universit\"{a}t, Max-von-Laue-Strasse 1, 60438 Frankfurt am Main, Germany}

\author{\firstname{M.} \surname{Kircher}}
\affiliation{Institut f\"{u}r Kernphysik, J. W. Goethe-Universit\"{a}t, Max-von-Laue-Strasse 1, 60438 Frankfurt am Main, Germany}

\author{\firstname{\\I.} \surname{Vela-P\'{e}rez}}
\affiliation{Institut f\"{u}r Kernphysik, J. W. Goethe-Universit\"{a}t, Max-von-Laue-Strasse 1, 60438 Frankfurt am Main, Germany}

\author{\firstname{J.} \surname{Siebert}}
\affiliation{Institut f\"{u}r Kernphysik, J. W. Goethe-Universit\"{a}t, Max-von-Laue-Strasse 1, 60438 Frankfurt am Main, Germany}

\author{\firstname{S.} \surname{Eckart}}
\affiliation{Institut f\"{u}r Kernphysik, J. W. Goethe-Universit\"{a}t, Max-von-Laue-Strasse 1, 60438 Frankfurt am Main, Germany}

\author{\firstname{S.} \surname{Grundmann}}
\affiliation{Institut f\"{u}r Kernphysik, J. W. Goethe-Universit\"{a}t, Max-von-Laue-Strasse 1, 60438 Frankfurt am Main, Germany}

\author{\firstname{G.} \surname{Kastirke}}
\affiliation{Institut f\"{u}r Kernphysik, J. W. Goethe-Universit\"{a}t, Max-von-Laue-Strasse 1, 60438 Frankfurt am Main, Germany}

\author{\firstname{M.} \surname{Waitz}}
\affiliation{Institut f\"{u}r Kernphysik, J. W. Goethe-Universit\"{a}t, Max-von-Laue-Strasse 1, 60438 Frankfurt am Main, Germany}

\author{\firstname{\\A.} \surname{Khan}}
\affiliation{Institut f\"{u}r Kernphysik, J. W. Goethe-Universit\"{a}t, Max-von-Laue-Strasse 1, 60438 Frankfurt am Main, Germany}

\author{\firstname{M.~S.} \surname{Sch\"{o}ffler}}
\affiliation{Institut f\"{u}r Kernphysik, J. W. Goethe-Universit\"{a}t, Max-von-Laue-Strasse 1, 60438 Frankfurt am Main, Germany}

\author{\firstname{F.} \surname{Trinter}}
\affiliation{Deutsches Elektronen-Synchrotron (DESY), FS-PE, Notkestrasse 85, 22607 Hamburg, Germany}
\affiliation{Fritz-Haber-Institut der Max-Planck-Gesellschaft, Molecular Physics, Faradayweg 4, 14195 Berlin, Germany}

\author{\firstname{R.} \surname{D\"{o}rner}}
\affiliation{Institut f\"{u}r Kernphysik, J. W. Goethe-Universit\"{a}t, Max-von-Laue-Strasse 1, 60438 Frankfurt am Main, Germany}

\author{\firstname{T.} \surname{Jahnke}}\email{jahnke@atom.uni-frankfurt.de}
\affiliation{Institut f\"{u}r Kernphysik, J. W. Goethe-Universit\"{a}t, Max-von-Laue-Strasse 1, 60438 Frankfurt am Main, Germany}

\author{\firstname{Ph.~V.} \surname{Demekhin}}\email{demekhin@physik.uni-kassel.de}
\affiliation{Institut f\"{u}r Physik und CINSaT, Universit\"{a}t Kassel, Heinrich-Plett-Str. 40, 34132 Kassel, Germany}

\date{\today}

\begin{abstract}
We suggest that low energy electrons, released by resonant decay processes, experience substantial scattering on the electron density of excited electrons, which remain a spectator during the decay. As a result, the angular emission distribution is altered significantly. This effect is expected to be a common feature of low energy secondary electron emission. In this letter, we exemplify our idea by examining the spectator resonant interatomic Coulombic decay (sRICD) of Ne dimers. Our theoretical predictions are confirmed by a corresponding coincidence experiment.
\end{abstract}

%\keywords{Photon interactions with molecules, Auger effect (including Coster-Kronig transitions), Photoelectron spectra}

\maketitle

The emission of secondary electrons after ionization or excitation of atoms and molecules has been vastly investigated since its discovery in 1905 by Pierre Auger. Such electronic decay processes provide unique information on electron-electron (configuration) interaction effects in matter. Auger decays can be grouped into two classes: so-called {\it participator} decays are cases, where the initially excited electron is actively participating in the decay by either being emitted or being the electron that fills a vacancy in an inner shell. In contrast, in {\it spectator} decays the initially excited electron does not participate in the decay but remains in its excited state acting simply as a spectator to the decay. It is commonly accepted that an electronic decay of an ionized or excited atom or molecule can be described in good approximation independently of the initial excitation step. As a consequence, for instance, an electron emitted by an Auger decay after photoionization, does not depend on the polarization properties of the absorbed photon \cite{Kuznetsov96}. This approximation is commonly known as the {\it two-step model} \cite{2ST}. For a resonant Auger decay  \cite{RA1,RA2}, this approximation is particularly valid if the excited electron is only witnessing the decay process as a spectator. However, several works have shown, that in special cases, the two-step approximation can break down \cite{Guillemin01prl,Weber03prl,Lagutin03,Liu08}.

In the present article we discuss a scenario of a breakdown of the two-step model which is not connected to specific, rare cases in nature, but is expected to occur very generally as soon as the electron emitted by the decay is of low kinetic energy. In this case, the Coulomb repulsion between the outgoing free electron and the excited bound electron may influence the emission direction of the former. Accordingly, for low energy electrons even a spectator electron is expected to influence the emission dynamics, as the wave packet of the slow secondary  electron will be scattered by the density of that spectator electron when escaping the system. Such final-state scattering effects, in general, should depend on the spatial symmetry of the excited electron, and information on the polarization of the exciting photon, which is imprinted in the symmetry of the spectator electron, is (in contrast to expectations from the two-step model) transferred to the secondary electron. Please note, that the effect discussed here is very different from the so-called post collision interaction \cite{PCI}, in which a high-energy Auger electron exchanges energy with a somewhat slower photoelectron.

A huge class of decay processes where this effect can be expected to occur routinely is interatomic (or intermolecular) Coulombic decay (ICD). Being predicted theoretically in 1997 \cite{Cederbaum97prlicd} and verified a few years later experimentally \cite{Marburger03,Jahnke04,Ohrwall04}, ICD and related processes have become a well-established and rapidly-growing  field of research (see, e.g., review articles \cite{ICD1,ICD2,ICD3,ICD4}). In general, ICD occurs in loosely bound matter as, e.g., van der Waals bound clusters or compounds bound by hydrogen bonds. In such systems, the energy released by a non-local electronic de-excitation of an atom or molecule is transferred to ionize a neighboring atom or molecule of the compound. Importantly, low energy electrons are typically emitted as a result of the interatomic (or intermolecular) decay process \cite{low1,low2,low3}. It has been demonstrated, that ICD can occur after a manifold of different excitation schemes.

Of specific interest in the present context is the so-called resonant interatomic Coulombic decay (RICD) \cite{RICD1,RICD2,RICD3,RICD4,RICD5,RICD6,RICD7,RICD8,RICD9}. In neon dimers, for example, an inner-valence Rydberg excitation (Ne$^\ast(2s^{-1}np)$Ne) can decay via the following three competing mechanisms \cite{RICD4}: (i) by autoionization (AI) of the Rydberg state, which is a purely atomic decay ionizing the initially  exited site of the dimer; (ii) by participator resonant ICD (pRICD), in which the $np$ Rydberg electron fills the $2s$ hole on the same site and the relaxation energy, transferred to the neighbor, is sufficient to ionize a $2p$ electron (i.e., the opposite site of the dimer becomes ionized in pRICD); and, finally, (iii) by spectator resonant ICD (sRICD), where the ICD process takes place in the presence of the excited Rydberg electron and where the initially excited atom remains excited and the opposite site becomes ionized. The two steps of sRICD in Ne dimers are as follows:
\begin{align}
\rm{Ne}_2 + \hbar\omega \rightarrow \rm{Ne}^*(2\textit{s}^{-1}\textit{np})\rm{Ne}~~~~~~~~~~~~~~~~~~~~~~~~ \nonumber \\
\rightarrow \rm{Ne}^*(2\textit{p}^{-1}\textit{np}) + \rm{Ne}^+(2\textit{p}^{-1}) + e_{ICD} \label{eq_final}
\end{align}

Due to symmetry, the AI and pRICD processes populate the same final states of the singly-ionized dimer and, thus, cannot be distinguished. As found by theoretical work \cite{RICD4}, the AI is by far dominant over pRICD. On the one hand, for highly-excited $np$  electrons, the sRICD rates converge to that of ICD of inner-valence ionized Ne dimers. Since both AI and pRICD processes involve Rydberg electrons, their decay rates fall rapidly and finally vanish  with increasing quantum number $n$. As demonstrated  in Ref.~\cite{RICD4}, for $3p$ excitation AI is the dominant relaxation pathway, and the sRICD process is almost two orders of magnitude weaker.  For $4p$ excitation, sRICD becomes similarly probable, and for $5p$ excitation even dominant over AI. The sRICD in Ne dimers has already been examined in a pioneering experiment by Aoto {\it et al.} \cite{RICD2}, where the angular resolved ion yield measurements allowed to distinguish the almost degenerate  excited states  $2\sigma^{-1}_{g/u} n\ell(\sigma/\pi)_{u/g}[\,^1\Sigma/ \Pi_u]$ of Ne$^\ast$Ne by their $\Sigma$ or $\Pi$, but not by $g$ or $u$ symmetry. Here we demonstrate the modification of the slow ICD electron angular emission  distribution by the spectator Rydberg electron in the course of sRICD in Ne dimers.

\begin{figure}
\includegraphics[scale=0.475]{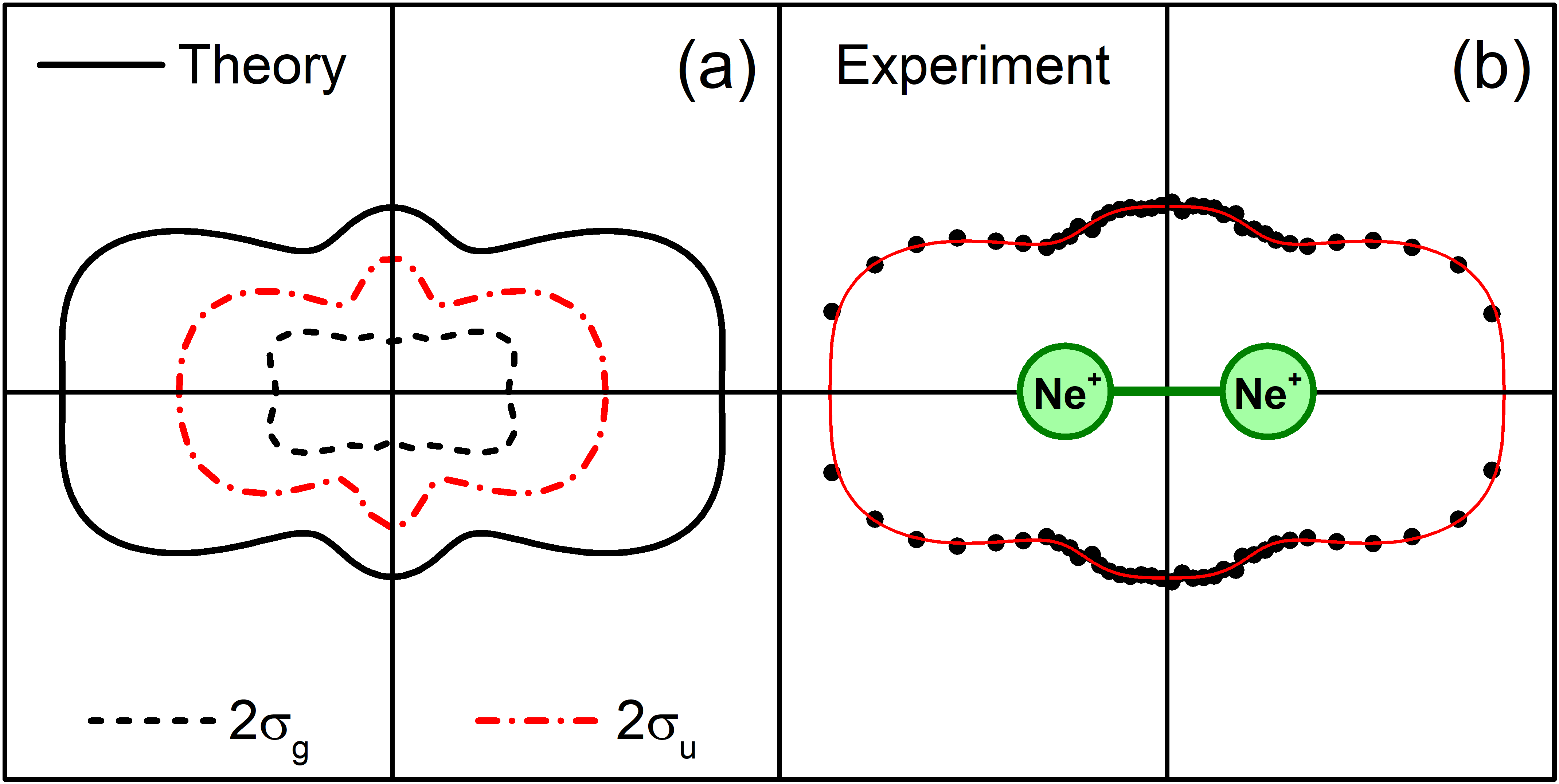}
\caption{Dimer frame angular emission distributions of ICD electrons emitted after $2s$ ionization. The dimer is oriented horizontally. {Panel (a):} Present calculations performed for an ICD electron energy of 0.65~eV, which corresponds to the maximum of the electron spectrum and to a decay at the equilibrium internuclear distance of 3.1~\AA~\cite{nICD2}. The total distribution is shown by the solid curve while contributions from the two initial states $2\sigma_{g/u}^{-1}$  are depicted by the broken curves. {Panel (b):} Experimental results, integrating over all occurring ICD electron energies. The full line is a fit with Legendre polynomials to guide the eye. Note: The distribution is independent of the orientation of the polarization vector of the ionizing photons with respect to the dimer axis \cite{nICD1}. }\label{nICD}
\end{figure}

In order to make a first estimate of the magnitude of the proposed effect, we have  examined the sRICD process (\ref{eq_final}) theoretically. Our calculations were performed employing the stationary Single Center (SC) method and code \cite{SC1,SC2}, which already provided an accurate description of angular resolved photoionization and decay spectra of diatomic molecules \cite{dia1,dia2,dia3} and weakly bound dimers \cite{RICD9,dia4} in the past. The transition amplitudes were computed within the frozen core Hartree-Fock approximation at different internuclear distances. The SC expansion of all occupied orbitals of Ne$_2$  with respect to the geometrical center of the dimer was restricted to partial harmonics with $\ell \leq 99$ and for the excited or ionized electron to partial waves with $\ell \leq 29$.

The computed partial transition amplitudes, describing the emission of electron waves with a fixed projection $m$ of the orbital angular momentum $\ell$ on the dimer axis, were used to obtain the electron angular emission distribution in the dimer frame. The derived working equations were tested by reproducing experimental \cite{nICD1} dimer-frame angular distribution of electrons emitted by ICD after inner-valence ionization of Ne dimers \cite{Jahnke04,nICD2}. Results of this test are depicted in Fig.~\ref{nICD}. The total electron angular distribution shown in Fig.~\ref{nICD}a (solid curve) consists of partial contributions from the two inner-valence ionized initial states ($g/u$) of the decay (broken curves). Each of these spectra include partial contributions from nine singlet and nine triplet two-site dicationic final states of ICD (not shown here for brevity). The agreement between our calculations and the experimental results shown in Fig.~\ref{nICD}b is very good, suggesting an appropriate modeling of the ICD process.

An accurate theoretical description of the sRICD process (\ref{eq_final}) requires precise potential energy curves of the excited initial and the singly-ionized and excited final states, as well as decay transition rates between them \cite{RICD4}. Subsequent nuclear dynamics calculations performed with the help of these data would provide the distribution of kinetic energies of the emitted sRICD electrons. After performing the aforementioned calculations, furthermore, the angular distribution of the sRICD electrons can be extracted. In order to estimate the magnitude of the proposed effect, we employed here a simplified one-particle approximation, in which a slow outgoing sRICD electron experiences, in addition to the potential of the final dicationic states, a potential generated by the spectator electron. As will become evident below, even this simplified model yields qualitatively appropriate dimer frame electron angular distributions (even though not on the level of those shown in Fig.~\ref{nICD}).

In order to model this situation theoretically, we have computed the wave functions of the excited $5p\sigma_{g/u}$ and $5p\pi_{g/u}$ Rydberg electrons, as described in our previous works \cite{RICD9,exc1,exc2,exc3,exc4}. The calculations were performed by employing the SC method at the equilibrium internuclear distance of 3.1~\AA~using the  potentials generated by the inner-valence ionized states $2\sigma_{g/u}^{-1}$. In the next step, the field generated by each spectator electron was added to the potential produced by each final two-site dicationic state, and the partial waves of a sRICD electron of 0.65 eV kinetic energy were computed (i.e., we used here the same kinetic energy as for the ICD electron in Fig.~\ref{nICD}). This simplified one-particle model does not include influences of the excited electron on the ionic core, which, in turn, affect the potential energy curves of the initial and final states of the decay. While lacking these features, the present model accurately describes the impact of multiple scattering of the outgoing electron and provides a very good estimate of the effects studied here.

In the final step, the wave functions of the sRICD electrons computed in the presence of excited states were used to calculate the respective decay transition matrix elements into all possible final doublet states and, additionally, to obtain the angular distribution of the emitted electrons in the frame of the dimer. All results were averaged over the almost degenerate electronic states of $g$ and $u$ symmetry, which cannot be  resolved in  experiments \cite{RICD2}. It should be stressed, that in the independent particle approximation, the matrix elements of sRICD do not involve radial parts of the excited electron and thus coincide with those of the ICD after inner-valence ionization. However, the information on the excitation is, in our case, imprinted in the wave functions of the low energy electrons, which do enter the decay transition matrix element.

\begin{figure}
\includegraphics[scale=0.475]{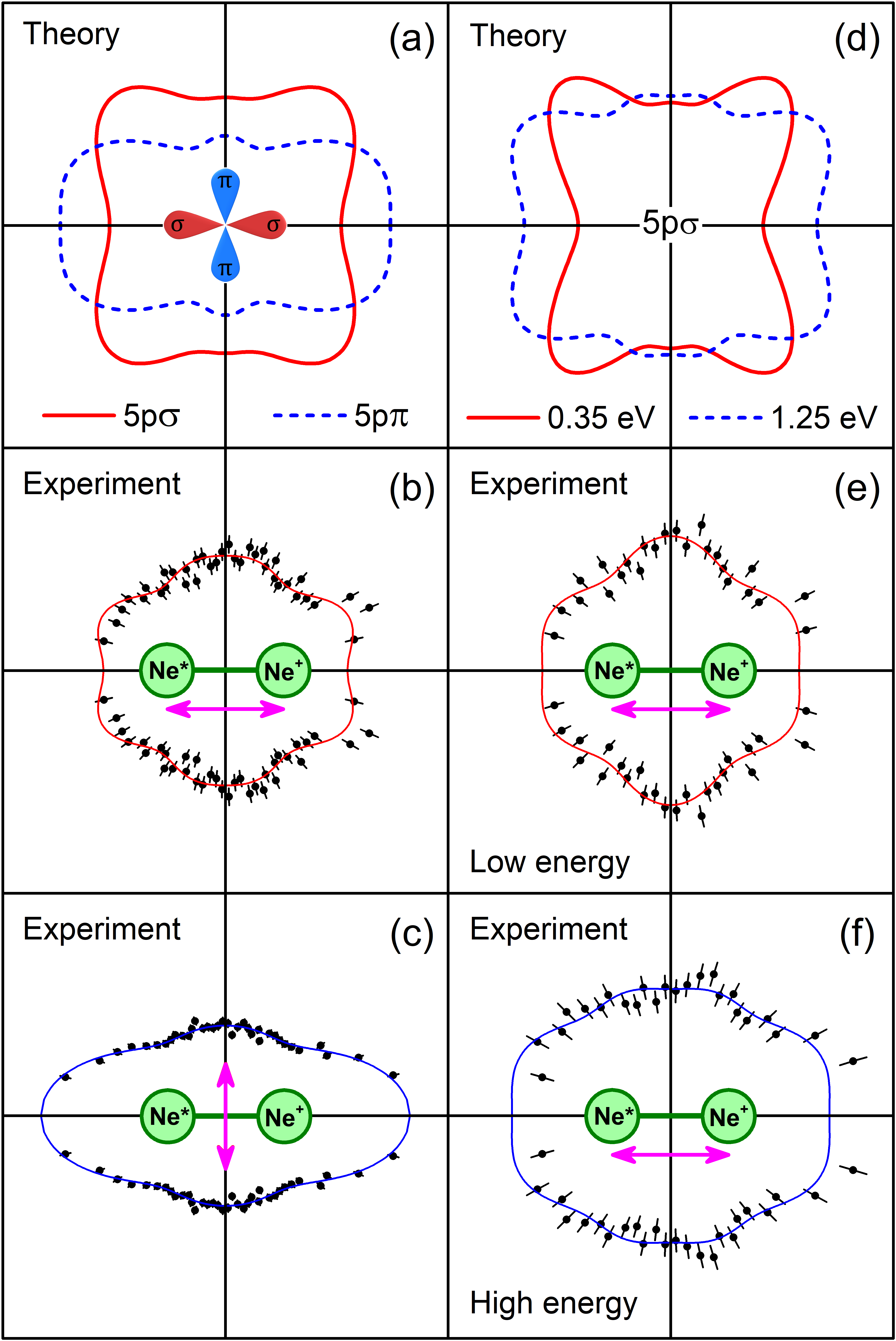}
\caption{Theoretical and experimental electron angular emission distributions of sRICD electrons in the dimer frame. Panel (a): the $5p\sigma$ and  $5p\pi$ spectator electrons are included in the potential for the calculation of  the continuum sRICD electron waves (see legends).  Panels (b) and (c): the dimer is oriented in parallel  or perpendicularly  to the polarization direction of the linearly polarized light used for the excitation (shown by double-arrows). Panel (d): calculations for the $5p\sigma$ state and two different kinetic energies of the sRICD electron (see legends).  Panels (e) and (f): measurements for the lower and higher kinetic energies of the sRICD electron and horizontal orientation of the electric field vector. Data in all panels include contributions from almost degenerate states of $g$ and $u$ symmetry.}\label{Results}
\end{figure}

The results of the theoretical modeling performed are shown in Fig.~\ref{Results}a. One can see from this figure, that the emission distributions obtained for $5p\sigma$ and $5p\pi$ excitations differ dramatically. The effect illustrated in Fig.~\ref{Results}a can be understood very intuitively. When escaping the dimer, the low energy electron is trying to {\it{avoid}} the spectator electron. Its density is sketched in Fig.~\ref{Results}a  for clarity. In the case of a $5p\sigma$ excitation, the additional electron density of the spectator is pointing along the dimer axis, which results in the suppression of the emission of sRICD electrons in this direction (as, for example, compared to the ICD electron emission pattern in Fig.~\ref{nICD}). In case of a $5p\pi$ excitation, the additional electron density is located perpendicularly to the dimer, and the low energy sRICD electrons are preferably emitted along the dimer axis.

If this intuitive picture is valid, it can be expected, that the observed effect depends strongly on the kinetic energy of the emitted electron. In order to check this expectation, we made calculations for the $5p\sigma$ excitation choosing somewhat lower and higher electron kinetic energies. The results obtained are depicted in Fig.~\ref{Results}d. As one can see, the scattering effects are indeed considerably larger for slower sRICD electrons, while the faster electrons have sufficient kinetic energy to penetrate through the electron density of the $5p\sigma$ spectator electron. The electron energies of 0.35~eV and 1.25~eV, selected for these simulations, correspond to distinct internuclear distances of 2.98~\AA~and 3.6~\AA~in the ICD process \cite{nICD2}. However, the observed changes cannot be related to an effect due to different internuclear distances, since the distributions of the ICD electrons emitted with different kinetic energies after inner-valence ionization (i.e., emitted at different internuclear distances) do barely differ as demonstrated in Ref.~\cite{nICD1}.

Figures~\ref{Results}b and \ref{Results}c show the corresponding results of our experiment on sRICD after $2s\to5p$ excitation of Ne dimers. Both figures confirm the theoretical predictions and show the same suppression of the electron emission along (Fig.~\ref{Results}b) and  perpendicular (Fig.~\ref{Results}c) to the dimer axis. Even the dependency on the kinetic energy of the emitted electron shown in Fig.~\ref{Results}d is qualitatively reproduced by the experiment (see Figs.~\ref{Results}e and \ref{Results}f).

The experiment has been performed at beamline UE112 of the Berlin synchrotron BESSY II \cite{Schiwietz_2015} by repeatedly scanning the photon energy across the maximum of the resonance located at 47.69~eV \cite{RICD2,NIST}. A Cold Target Recoil Ion Momentum Spectroscopy (COLTRIMS) setup \cite{COLTRIMS1,COLTRIMS2,Jahnke04JESRP} has been employed to measure the momenta of all charged particles created after the absorption of the photon in coincidence. A supersonic gas jet composed of 70\% Ne and 30\% He is crossed with the synchrotron beam at right angle yielding a well-defined reaction volume. By cooling the gas jet nozzle (with a diameter of $5~\rm{\mu m}$) to $85~$K and using a driving pressure of 6.3~bar, a small fraction of  neon dimers is created in the supersonic expansion while larger clusters are still absent. Static electric and magnetic fields are used to guide charged fragments to two time- and position-sensitive multi-channel plate detectors with delayline position readout \cite{Jagutzki02}.  The ion arm of the COLTRIMS analyzer consisted of a single acceleration region with a length of $7$~cm. The electron arm incorporated a time-focusing geometry with a $6$~cm long acceleration region (3.6~V/cm) followed by a $12$~cm long drift region. A homogeneous B-field, parallel to the electric field, of 2.55~G was used to achieve a detection solid angle of $4\pi$ for electrons with energies of up to 6~eV.

The trajectories of the emitted particles inside the COLTRIMS analyzer are reconstructed from the measured positions of impact and respective times-of-flight. Thereby, the initial vector momenta of the particles are deduced and all derived quantities (as, e.g., their kinetic energies and emission angles) are obtained. The total energy released by the decay (\ref{eq_final}) is given by: $E_{tot}=E_{\rm{Ne}^\ast(2s^{-1}5p)}-E_{\rm{Ne}^\ast(2p^{-1}5p)}-E_{\rm{Ne^+}(2p^{-1})}$. This energy is shared between the emitted sRICD electron and the kinetic energy of the fragments in the center of mass frame (KER). Therefore, valid events of sRICD have been identified as the measured total energy $E_{tot}=KER+\varepsilon_{ICD}$ is within the interval of 4.75 -- 5.75~eV \cite{NIST}, thus removing monomer events and background. The coincident detection of the electron and the emitted ion provides further information: Firstly, the orientation of the dimer in the laboratory frame can be deduced. This is possible if the break-up of the dimer happens rapidly after the decay and, accordingly, the dimer does not have time to rotate \cite{Weber01jpb}. In that case the orientation of the dimer in the laboratory frame equals in a very good approximation the emission direction of the Ne$^+$ ion. Secondly, the electron angular distribution in the body fixed frame of the dimer can be reconstructed as the relative emission angle between the electron and the ion is obtained from the coincidence measurement \cite{Jahnke2002CON2}.

In summary, we examined the spectator resonant ICD after inner-valence excitation of Ne dimers into the $5p$ state. We observe a strong dependency of the ICD electron angular emission distribution on the relative orientation of the dimer with respect to the polarization axis of the exciting photons. While this seemingly contradicts the well-established  two-step model of excitation and decay, it turns out, that it can be fully described by a theory which relies on this approximation. According to the present theory, this effect occurs due to a scattering of the low energy ICD electron at the excited anisotropic Rydberg electron. Thereby, the dependency of the angular emission distributions on the polarization direction, predicted theoretically and verified experimentally, is due to a selective excitation into  the $5p$ states of either $\sigma$  or  $\pi$ symmetry, which have very different spatial density distributions. The observed effect is expected to always take place whenever a low kinetic energy electron is emitted in the presence of an excited spectator electron. The observations reported here can thus be considered relevant to an extended class of decay processes and should be taken into account when interpreting theoretical or experimental investigations of such processes.

\begin{acknowledgments}
This work was supported by the Deutsche Forschungsgemeinschaft (DFG) within the research unit FOR 1789. We would like to express our gratitude towards the staff of the Berlin synchrotron BESSY II for their support. We thank HZB for the allocation of synchrotron radiation beamtime, and we thankfully acknowledge the financial support by HZB.
\end{acknowledgments}

\end{document}